\documentclass[12pt]{iopart}
\usepackage{verbatim} 
\usepackage{graphicx}
\usepackage{kevin}
\begin{document}

\title[Role of quantum fluctuations in a system with strong fields]{Role of quantum fluctuations in a system with strong fields}

\author{Kevin Dusling}

\address{North Carolina State University
Box \#8202
Raleigh, NC 27695}
\ead{kevin\_dusling@ncsu.edu}
\begin{abstract}
In this work we study how quantum fluctuations modify the quantum evolution of an initially classical field theory.  We consider a scalar $\phi^4$ theory coupled to an external source as a toy model for the Color Glass Condensate description of the early time dynamics of heavy--ion collisions.
We demonstrate that quantum fluctuations considerably modify the time evolution driving the system to evolve in accordance with ideal hydrodynamics.  We attempt to understand the mechanism behind this relaxation to ideal hydrodynamics by using modified initial spectra and studying the particle content of the theory. 
\end{abstract}

\maketitle

\section{Introduction}

One of the outstanding theoretical problems in heavy--ion physics is a first principles understanding of the isotropization and thermalization of the matter produced in collision.  The fact that the system is nearly thermal and isotropic at early times has been deduced from hydrodynamic model fits \cite{Romat1,SongH1,LuzumR1,DusliMT1,Song:2010mg,Song:2011hk} to the measured spectra and elliptic flow~\cite{Adamsa3,Adcoxa1,Arsena2,Backa2}.  Estimates of the relaxation time range from $\tau_{\rm relax}\sim 0.5$--$2$ fm \cite{LuzumR1} which is hard to accommodate within a simple picture of interacting quasi--particles. 

However, a quasi--particle description is not essential to thermalization and in this work we will demonstrate that an initially strong classical field undergoing quantum evolution may evolve in accordance with ideal hydrodynamics.  We will show that the presence of secular divergences (modes whose occupation number grows with time) become semi--classical on relatively short time scales and must be resumed to all orders in a standard perturbative expansion.

While this works focuses on scalar $\phi^4$ theory it is suggested that similar mechanisms may be at work within the framework of the Color Glass Condensate (CGC) description \cite{IancuV1,IancuLM3,GelisIJV1,GriboLR1,MuellQ1,McLerV1,McLerV2,McLerV3} of high energy nuclei.  Work in applying the techniques shown here to the case of classical Yang--Mills is currently in progress \cite{DusliGSV1}.  The goal of this work is to understand the role secular divergences play in modifying the time evolution of the classical field.  We will demonstrate that the quantum evolution of a scalar field evolves in accordance with ideal hydrodynamics.  Finally, we will speculate on the mechanism behind this relaxation by looking at modified spectra and the time  evolution of the number density.  

This work is largely based off the first paper on this topic \cite{Dusling:2010rm} which we refer the reader to for more details on the resummation scheme.  It is worth pointing out that the resummation of secular diverges is qualitatively similar to the resummation of leading logarithms $(g^2 \ln(1/x_{1,2}))^n$ of the incoming partons' momentum fractions required for the computation of inclusive quantities at leading log order \cite{GelisLV3,GelisLV4,GelisLV5}.  These results have proven to be valuable in a quantitative understanding of the near side angular correlations observed in nucleus--nucleus \cite{Alver:2009id,Dusling:2009ni} and proton--proton \cite{Khachatryan:2010gv,Dumitru:2010iy} collisions.  We would like to point out that a considerable amount of work has been done for theories similar to $\phi^4$ in the context of reheating after inflation \cite{ProkoR1,Frolo1,FeldeT1}.  In addition, considerable progress has been made on the thermalization problem in the context of Heavy--Ion collisions through the use of N--particle irreducible effective actions \cite{Berges:2004yj,Berges:2008wm,Aarts:2001yn,Aarts:2001qa,Berges:2002cz,Aarts:2000mg}.

\section{The Model}

The CGC inspired scalar theory model has the Lagrangian
\beqa
\mathcal{L}=\frac{1}{2}\left(\partial_\mu \phi\right)\left(\partial^\mu\phi\right) - V(\phi) + J\phi
\eeqa
where the interaction potential is
\beqa
V(\phi)=\frac{g^2}{4!}\phi^4
\eeqa
and J is an external source which mimics the large $x$ color charges of the incoming nuclei.  Since the external current vanishes after the collision takes place we take our source to be nonvanishing for $x^0 < 0$ only,
\beqa
J(x)\sim \theta(-x^0)\frac{Q^3}{g}\,.
\eeqa

The role of the external source is to initialize a classical field (having occupation number $\sim 1/g^2$) which evolves solely via their self--interactions at $x^0\geq 0$.  This source term also brings an external scale into the problem.  Since we assume the source is turned on adiabatically from $x^0\to -\infty$ we can consider the evolution of the classical field as an initial value problem at $x^0=0$ with $\phi(x^0=0)\sim Q/g$ and $\dot{\phi}(x^0=0)=0$. 

Throughout this work we will use the same model parameters.  In order to avoid confusion we now state these parameters once and for all.  For both the homogeneous and non--homogeneous systems we take $\phi_0=12, \dot{\phi}_0=0$ and $g=0.5$ (a very weak coupling considering the factor of $4!$ in front of the potential).  For the case of the 3D simulations we employ a $12^3$ lattice with a volume of $12^3$.  This lattice size will have a momentum cutoff of $k_{max}\approx 5.44$.  For these model parameters the resonance mode exists between $3\lsim k_{res} \lsim 3.22$ at $t=0$.  The zero--mode has an effective mass $m^2\equiv (g\phi_0)^2/2 = 18$ and a period of oscillator of $T\approx 3$.

\section{Homogeneous System \& Homogeneous Fluctuations}

In this section we consider a classical background field that is homogeneous in all space and undergoes quantum evolution with space--independent fluctuations ({\em i.e.} zero--mode fluctuations).  While highly unrealistic this simple toy model will allow us to see how the mechanism of phase decoherence leads to ideal hydrodynamic evolution.  The Lagrangian for a uniform non--expanding scalar theory is
\beqa
\mathcal{L}=\frac{1}{2}\dot{\phi}^2-V(\phi)\,,
\eeqa
where $\dot{\phi}=d\phi/dt$.  The classical evolution can be found in closed form.  Since the energy $\mathcal{H}=\dot{\phi}^2/2+V(\phi)$ remains constant throughout the evolution we can write
\beqa
\frac{1}{2}\dot{\phi}^2=E_0-V(\phi)\,,
\label{eq:E0}
\eeqa
where $E_0$ is the initial energy of the system which is determined by the initial condition of our classical field
\beqa
E_0=\frac{1}{2}\dot{\phi}_0^2+\frac{g^2}{4!}\phi_0^4\,.
\eeqa
where $\phi_0=\phi(t=t_0)$.  Equation \ref{eq:E0} can be integrated to obtain
\beqa
t-t_0=\frac{1}{\sqrt{2}}\int_{\phi_0}^{\phi(t)}\frac{d\psi}{\sqrt{E_0-V(\psi)}}\,.
\eeqa
At this point it will be useful to introduce some notation.  Let us define
\beqa
\epsilon^2\equiv g^2/4!\,,
\eeqa  
and make the change of variables $\sqrt{\epsilon}\psi=-E_0^{1/4}\cos\theta$.  We are also free to set $t_0=0$ for the non--expanding case and we find
\beqa
t=\frac{1}{2\sqrt{\epsilon}E_0^{1/4}}\int_{\theta_0}^{\theta(t)} \frac{d\phi}{\sqrt{1-\frac{1}{2}\sin^2\phi}}
\eeqa
where
\beqa
\theta(t)&=&\cos^{-1}\left(\frac{\sqrt{\epsilon}\phi(t)}{E_0^{1/4}}\right)\nonumber\\
\theta_0&=&\cos^{-1}\left(\frac{\sqrt{\epsilon}\phi_0}{E_0^{1/4}}\right)
\eeqa
The above integral equation can be solved for $\phi(t)$ in terms of the Jacobi elliptic function of the first kind having elliptic modulus $1/2$.
\beqa
\phi(t)=\frac{E_0^{1/4}}{\sqrt{\epsilon}}\textrm{cn}_{1/2}\left[2\sqrt{\epsilon}E_0^{1/4}t-\textrm{F}_{1/2}(\theta_0)\right]
\label{eq:phizero}
\eeqa
where $\textrm{F}_{1/2}(\theta_0)$ is the incomplete elliptic integral of the first kind of modulus $1/2$.  The above result is periodic with period
\beqa
T=\frac{2}{\sqrt{\epsilon} E_0^{1/4}}\textrm{K}(1/2)
\label{eq:T}
\eeqa
where $K(1/2)\approx 1.85407$ is the complete elliptic integral of the first kind. Notice that the period of oscillations depends on the initial conditions through $E_0$.  The fact that the period of oscillation depends on the initial condition is a signature of non--linear evolution and is crucial for phase decoherence. 

It is worth noting that to a very good approximation (within about 15\%) the above expression for $\phi$ can be approximated by
\beqa
\phi(t)\approx \frac{E_0^{1/4}}{\sqrt{\epsilon}}\cos\left[\frac{2\pi}{T}( t - \xi)\right]\,,
\eeqa
where $\xi$ is a phase set by the initial conditions
\beqa
\xi=\frac{\theta_0}{2\sqrt{\epsilon}E_0^{1/4}}\,.
\eeqa

\subsection{Stress--energy tensor}

With an analytic expression for $\phi(t)$ available we can now find analytic expressions for the stress energy tensor as well.  For the homogeneous non-expanding system there are two independent components of the stress--energy tensor
\beqa
T^{00}&=&\frac{1}{2}\dot{\phi}^2+V(\phi)\nonumber\\
T^{ij}&=&\delta^{ij}\left(\frac{1}{2}\dot{\phi}^2-V(\phi)\right)
\eeqa
with all other components vanishing.  Using the expressions derived in the previous section we find
\beqa
T^{00}&=&E_0\nonumber\\
T^{ij}&=&E_0\left[1-2\textrm{cn}^4_{1/2}\left(2\sqrt{\epsilon}E_0^{1/4}t-\textrm{F}_{1/2}(\theta_0)\right)\right]\nonumber\\
&\approx&E_0\left[1-2\cos^4\left(\frac{2\pi}{T}( t -   
\xi)\right)\right]
\eeqa

\begin{figure}
\begin{center}
\includegraphics[scale=.8]{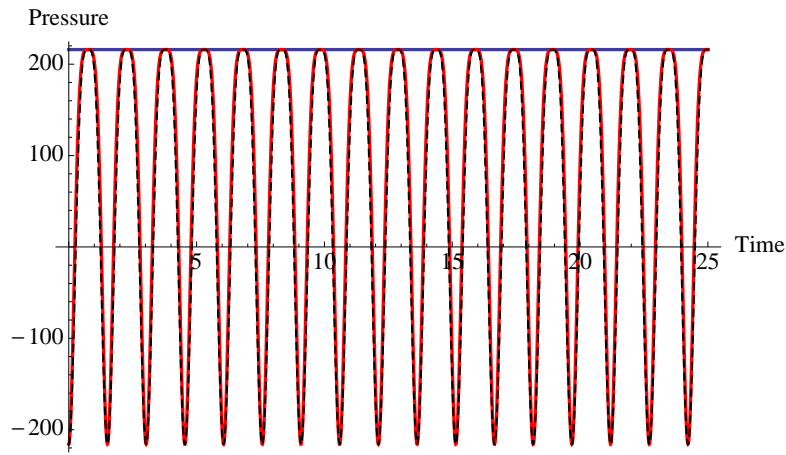}
\caption{$T^{00}$ and $T^{11}$ for a uniform non--expanding $\phi^4$ theory.  The dashed line uses the approximation for $T^{11}$ explained in the text.}
\label{fig:phi4_uniform}
\end{center}
\end{figure}

As an example, in Fig.~\ref{fig:phi4_uniform} we plot $T^{00}$ and $T^{11}$ as a function of time.  We also show the good agreement between the true solution and approximate form of $T^{11}$ in this figure.  Clearly, this LO result does not have a well defined equation of state. 

\subsection{Spectrum of fluctuations}

We now want to superimpose quantum fluctuations on top of our classical background field.  In this section we will consider the following toy model for the spectrum of fluctuations,
\beqa
\textrm{F}(a,\dot{a})=\delta(\dot{a})\frac{1}{\sqrt{2\pi\sigma^2}}\exp\left(-\frac{a^2}{2\sigma^2}\right)\,,
\eeqa
where $\sigma$ characterizes the variance of the zero--mode fluctuations.  For this toy model we will treat $\sigma$ as a free parameter.  It will be computed from first principles later on.  We should stress that this is a highly unrealistic model since we are ignoring any quantum fluctuation which are non-homogeneous in space.   

\begin{figure}
\begin{center}
\includegraphics[scale=.8]{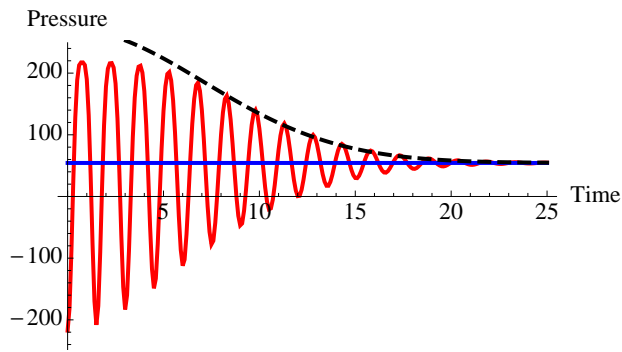}
\caption{$T^{00}/3$ and $T^{11}$ for a uniform non--expanding $\phi^4$ theory after averaging over a Gaussian distribution of fluctuations in $\phi_0$ having $\sigma=0.4$.  The dashed curve shows the envelope as given in the text.}
\label{fig:phi4_uniform_avg}
\end{center}
\end{figure}

The expectation value of an inclusive operator (such as the stress energy tensor) is defined as
\beqa
\langle\mathcal{O}\rangle=\int_{-\infty}^{+\infty}da\spc d\dot{a}\spc \textrm{F}(a,\dot{a}) \mathcal{O}_{\rm LO}(\phi_0+a,\dot{\phi}_0+\dot{a})
\eeqa
where $\mathcal{O}_{\rm LO}(\phi_0+a,\dot{\phi}_0+\dot{a})$ is the operator of interest computed at leading order with initial conditions shifted by $a$ and $\dot{a}$.  For this particular choice of fluctuations and using our approximate solutions for $\phi(t)$ found in the previous section, the integrals over $a$ and $\dot{a}$ can be done analytically.  The result is shown in fig.~\ref{fig:phi4_uniform_avg} for $\sigma=0.4$.  The analytic expression is not too enlightening.  It essentially consists of a number of terms having oscillations at different frequencies which die off exponentially at different rates.  But it is instructive to pull out the one term which dies off slowest.  Its envelope is given by 
\beqa
\sim e^{-2c^2g^2\sigma^2t^2}\,,
\eeqa
where we have defined the constant $c\equiv\frac{\pi}{K(1/2)\sqrt{4!}}\approx 0.3459$.  We can now identify a relaxation time
\beqa
\tau_{\rm relax}=\frac{1}{\sqrt{2}cg\sigma}\approx \frac{2}{g\sigma}\,.
\label{eq:tauhomo}
\eeqa
While this is a very unrealistic model it is nice that the above result could be derived analytically.  In the example shown in Fig.~\ref{fig:phi4_uniform_avg} the fluctuations are completely absent after $2\times \tau_{\rm relax}$.  In other words, by $t\sim 4/(g\sigma)\approx 20$ the system has a well defined equation of state ($\epsilon=3p$) and evolves in accordance with ideal hydrodynamics.

The mechanism behind the relaxation of the pressure is quantum decoherence which we now explain.  Each initial condition in our ensemble average is shifted by a random Gaussian variable ($\phi_0\to \phi_0+a$) and this corresponds to a shift in the initial energy $E_0$ of the system.  The time evolution of each individual system is periodic with a slightly different period of oscillation as given by Eq.~\ref{eq:T}.  When performing the ensemble average the differing periods of each system results in a phase decoherence forcing the pressure to relax to its equilibrium value.  Let us stress that this will not occur in a $\phi^2$ theory.  In this case the period of oscillation will {\em not} depend on the initial condition.

\section{Non--Homogeneous Fluctuations}

In the previous section we showed how a homogeneous system undergoing zero--mode fluctuations relaxes to a system evolving according to ideal hydrodynamics.  While the previous case is of pedagogical interest since it shows simply how the decoherence of the quantum field leads to the relaxation of the pressure it is highly unrealistic in that it does not include space dependent fluctuations.  

We now consider the same model in three dimensions including the space dependent fluctuations as predicted from quantum field theory.  In order to motivate the need for the resummation we first discuss the case of linearized perturbations.
 
\subsection{Linear perturbations}

\begin{figure}
\begin{center}
\includegraphics[scale=0.5]{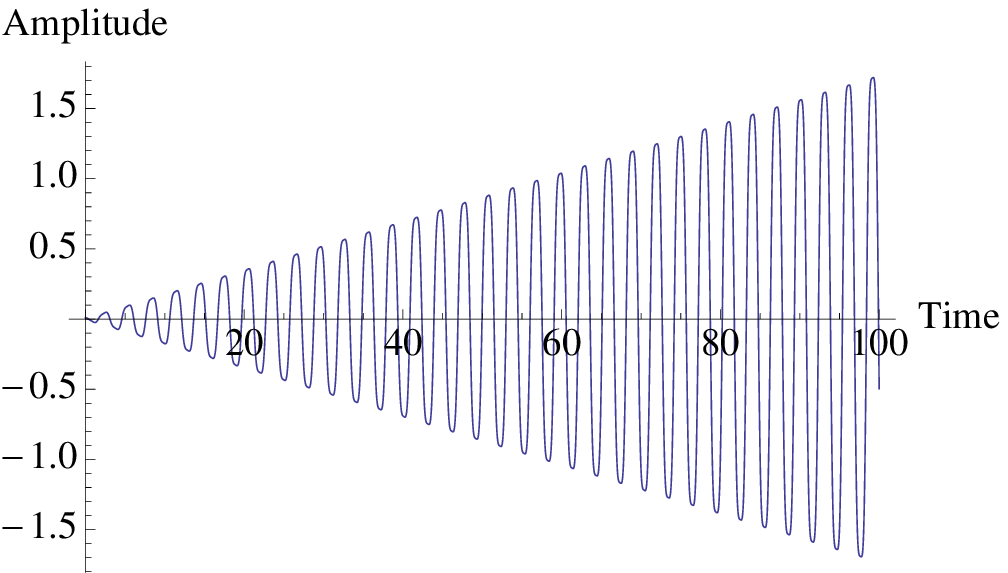}
\includegraphics[scale=0.5]{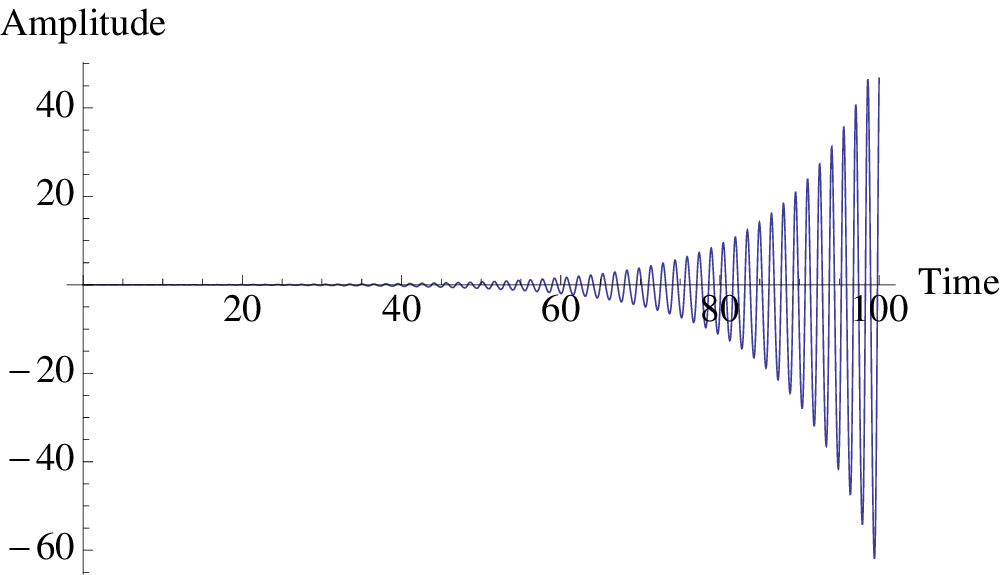}
\includegraphics[scale=0.5]{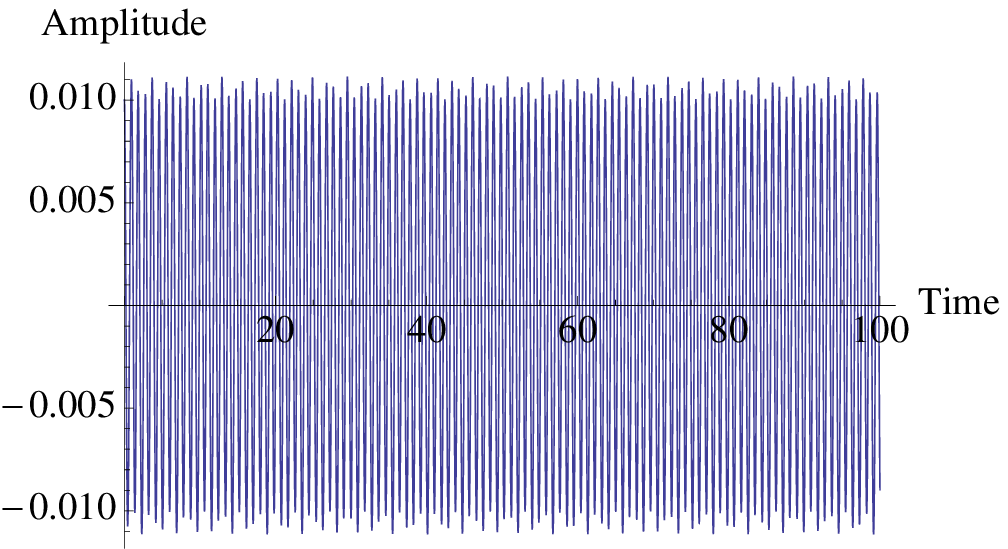}
\caption{Evolution of a linear perturbation on top of the background field.  Left: Linear growth of the ${\bf k}=0$ mode.  Middle: Exponential growth of a resonant mode ${\bf k}_{res}=0.5165g\phi_0$.  Right: Typical behavior of a stable perturbation ${\bf k}=2{\bf k}_{res}$.}
\label{fig:linear}
\end{center}
\end{figure}

In this section we now consider how a linearized perturbation evolves on top of the homogeneous background field.  We decompose the background field into a homogeneous part $\phi_{\bf k=0}$ and a small field perturbation $a(x)$. The equation of motion for the Fourier transform of our field perturbation $a(x)$ is 
\beqa
\ddot{a}_{\pm{\bf k}}+\left[{\bf k}^2+V^{\prime\prime}(\phi_{\bf k=0})\right]a_{\pm{\bf k}}=0\,.
\label{eq:linear}
\eeqa
In the above expression $\phi_{\bf k=0}$ is the zero--mode solution given by Eq.~\ref{eq:phizero}. We now numerically solve Eq.~\ref{eq:linear} in order to investigate how linear perturbations evolve on top of the background field.  In Fig.~\ref{fig:linear} we show how the amplitude of three ${\bf k}$ modes evolve when given an initial amplitude of $a_{\bf k}(t=0)=0.1$.  The first plot shows the zero mode whose amplitude grows linearly with time.  The second mode is taken from within the resonance band and it clearly grows exponentially with time.  The third mode shows the typical behavior of a high momentum mode (here shown for $k=2k_{res}$).  While the high momentum modes can be treated perturbatively as their amplitude does not grow with time the lower ${\bf k}$ modes lead to secular divergences.  Clearly, at times when $gt$ (for modes outside the resonance band) or $ge^{\mu t}$ (for resonance modes) become of $\mathcal{O}(1)$ a resummation becomes necessary.

\section{Results from the full fluctuation spectrum}

\subsection{Initial Condition}

As is clear from the previous discussion quantum fluctuations on top of the homogeneous background field will play an important role in the resulting dynamics.  The spectrum of these quantum fluctuations are derived from first principles.  In this case the classical field $\phi$ and its conjugate momentum $\pi\equiv \partial \mathcal{L}/\partial\dot{\phi} =\dot{\phi}$ are promoted to quantum operators $\hat{\phi}$ and $\hat{\pi}$ obeying the equal time commutation relations
\beqa
\left[\hat{\phi}({\bf x}),\hat{\pi}({\bf y})\right]=i\delta^3({\bf x}-{\bf y})
\eeqa
The field operators can be rewritten in terms of creation and annihilation operators
\beqa
\hat{\phi}(x)=\frac{1}{(2\pi)^{3/2}}\int \frac{d^3k}{\sqrt{2\omega_{\bf k}}}\left[ \hat{a}^\dagger_{\bf k}e^{ik_\mu x^\mu} + \hat{a}_{\bf k}e^{-ik_\mu x^\mu} \right]\,,
\eeqa
obeying
\beqa
\left[\hat{a}_{\bf k},\hat{a}^\dagger_{\bf p}\right]=\delta^3({\bf k}-{\bf p})\,.
\eeqa
Using the above mode decomposition one can easily show that the two-point correlation function in a homogeneous background field takes the form
\beqa
\langle \hat{\phi}({\bf x}) \hat{\phi}({\bf y}) \rangle &=& \frac{1}{(2\pi)^3}\int \frac{d^3k}{2\omega_{\bf k}} e^{i{\bf k}\cdot\left({\bf x}-{\bf y}\right)}\,,\\
\langle \hat{\pi}({\bf x}) \hat{\pi}({\bf y}) \rangle &=& \frac{1}{(2\pi)^3}\int \frac{d^3k}{2}\omega_{\bf k} e^{i{\bf k}\cdot\left({\bf x}-{\bf y}\right)}\,,
\eeqa
where $\omega_{\bf k}^2={\bf k}^2+m^2$.  Our semi--classical simulation will therefore consists of a Gaussian random field having power spectrum
\beqa
\mathcal{P}_{\phi}({\bf k})=\frac{1}{2(2\pi)^3 \omega_{\bf k}}
\eeqa
superimposed on top of the homogeneous background field.  The power spectrum as written above is UV divergent and this is regulated by the lattice spacing.  If we impose a momentum cutoff $\Lambda$ the energy density will contain terms that behave parametrically as $Q^4/g^2, Q^2\Lambda^2$ and $\Lambda^4$.  The $\Lambda^4$ is a pure vacuum contribution and can be computed by performing simulations with the source $J$ turned off which can then be subtracted from the corresponding result.  The $Q^2\Lambda^2$ terms in not renormalizable in the usual sense since it mixes diagrams having an arbitrarily high number of loops.  In practice, what is done, is to choose a cutoff that is sufficiently large in order to encompass the relevant physics ($\Lambda \gsim m$) but small enough to keep the cutoff--dependent terms negligible with respect to the classical contribution ($\Lambda \ll Q/\sqrt{g}$).

\subsection{Results}

\begin{figure}
\begin{center}
\includegraphics[scale=1.0]{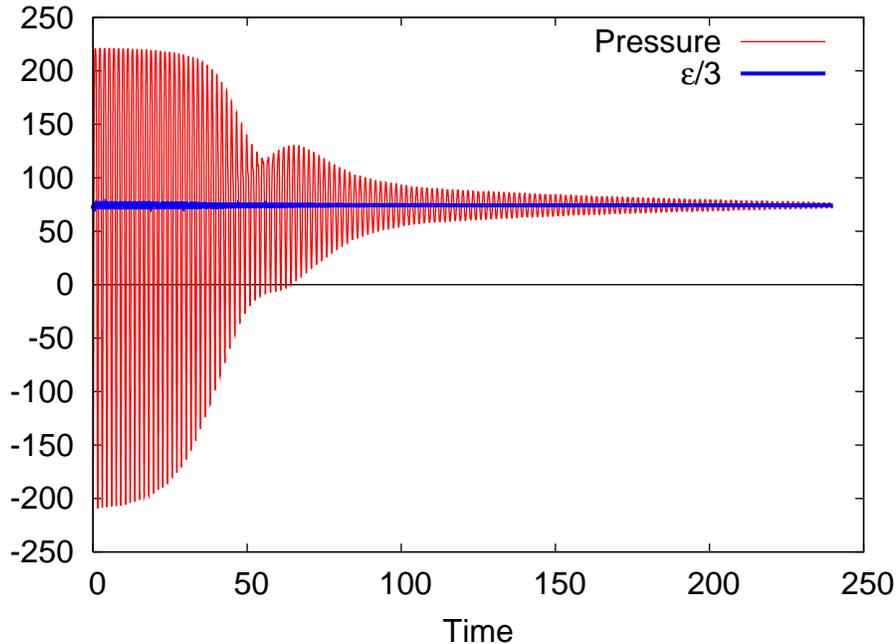}
\caption{Time evolution of the pressure averaged over an ensemble of 1000 configurations using the spectrum of fluctuations as given from quantum field theory.}
\label{fig:Full}
\end{center}
\end{figure}

Fig.~\ref{fig:Full} shows the pressure and energy density ($\epsilon/3$) as a function of time with an ensemble average of 1000 simulations.  The main conclusion of this paper is that the ensemble averaged pressure relaxes towards $\epsilon/3$ and therefore has a well--defined equation of state and evolves in accordance with ideal hydrodynamics.  We should stress that even though there exists a well--defined ({\em i.e.} time--independent) equation of state the system is not necessarily in thermal equilibrium as well will show.  This rapid establishment of an equation of state known as {\em prethermalization} has been studied in the context of a linear $\sigma$--model using the 2PI effective action in \cite{Berges:2004ce}.  This work found similar conclusions; there can be the rapid establishment of an equation of state via phase decoherence regardless if scattering processes can thermalize the system. 

  It is apparent from figure~\ref{fig:Full} that the time evolution evolves in two stages.  First, in the window $0\leq t\lsim 50$ the amplitude of the pressure oscillations decrease very quickly to moderate values.  Then from a time $t\sim 50$ and onwards there is a slight rebound and a gradual approach to complete relaxation.

\begin{figure}
\begin{center}
\includegraphics[scale=0.6]{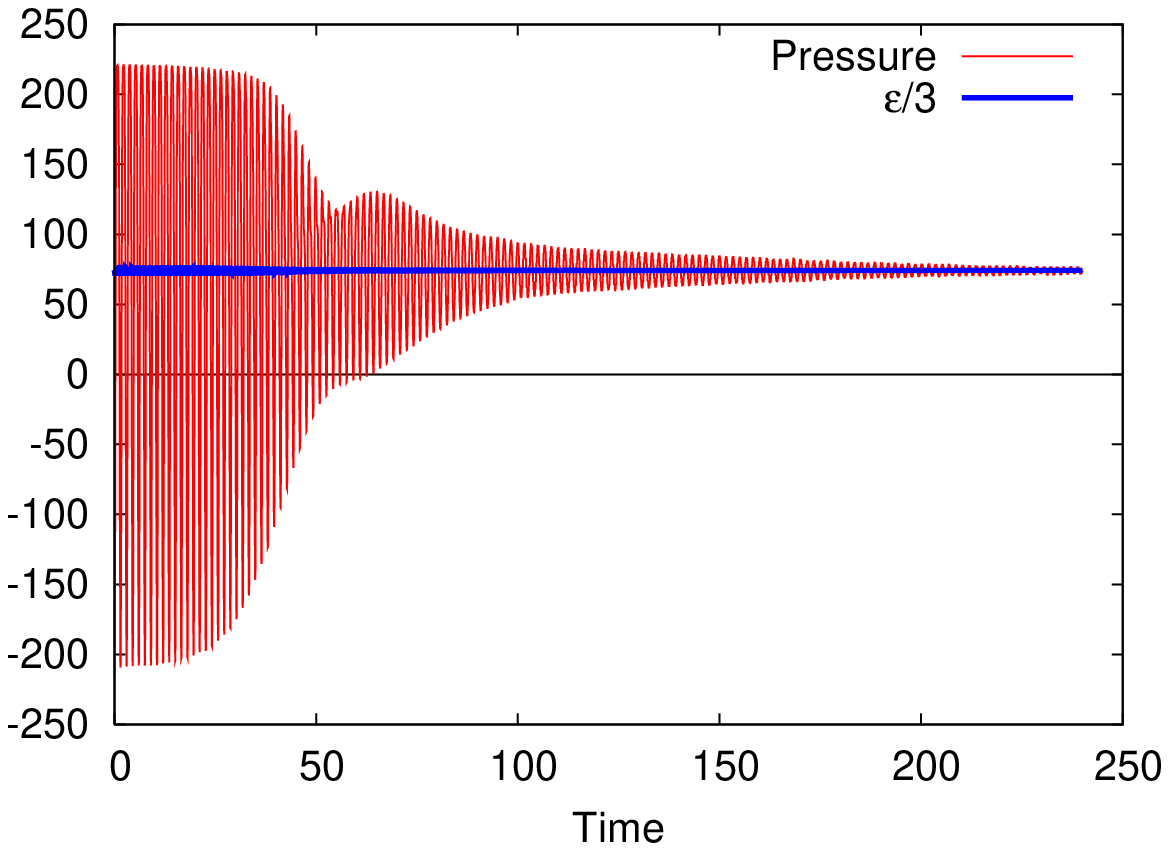}
\includegraphics[scale=0.6]{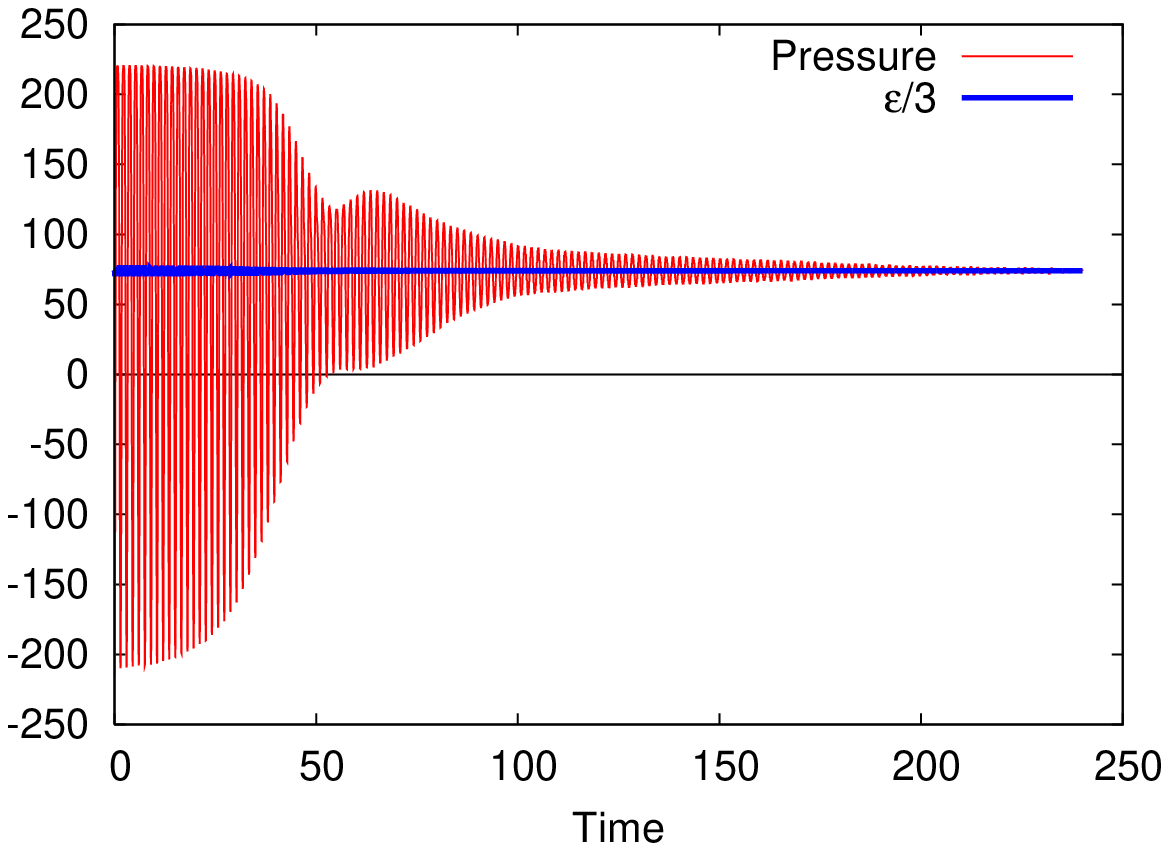}\\
\includegraphics[scale=0.6]{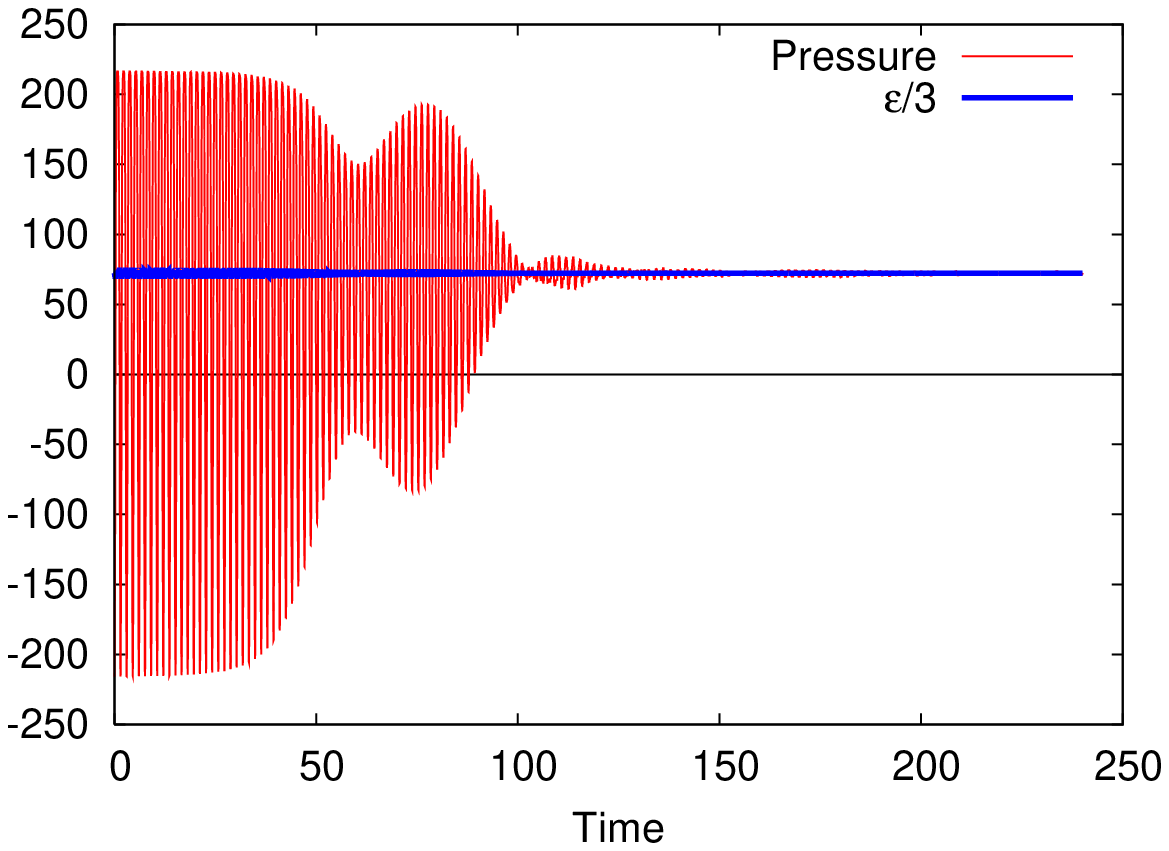}
\includegraphics[scale=0.6]{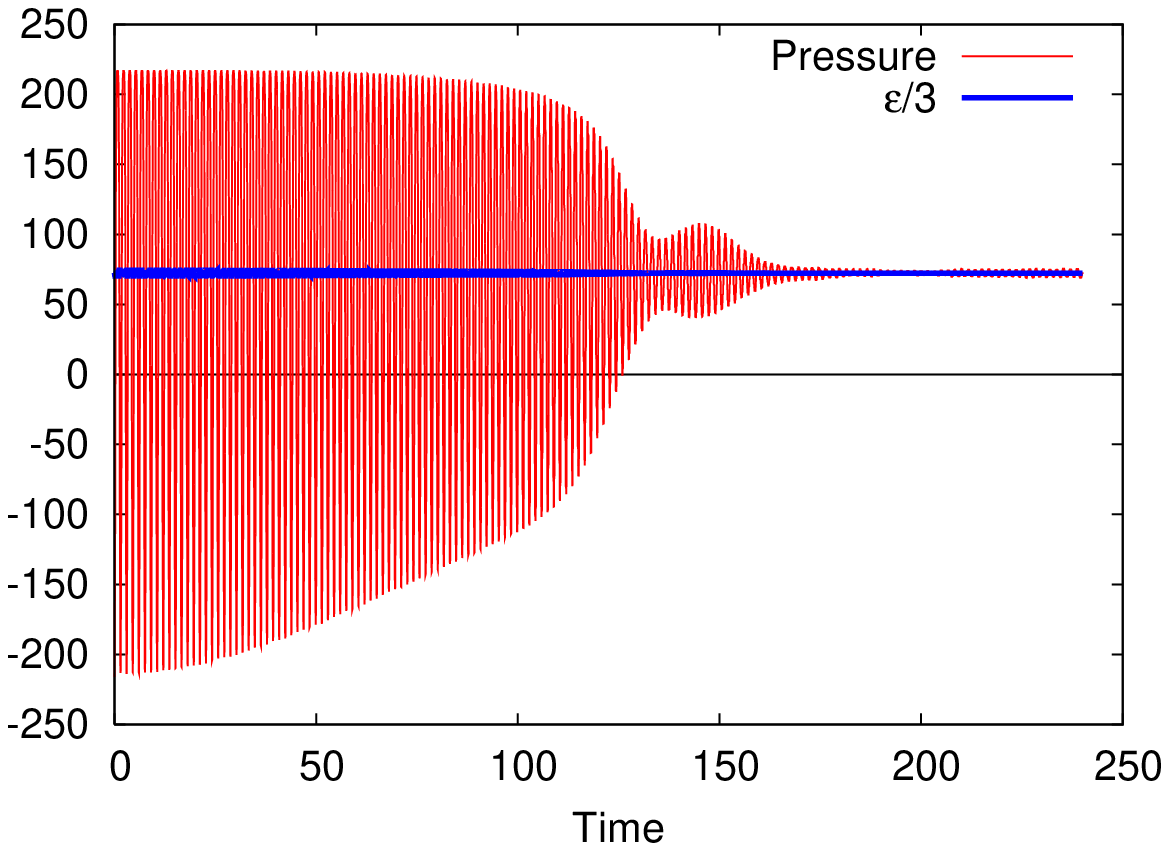}
\caption{Evolution of the ensemble averaged pressure for various {\em incorrect} spectrum of fluctuations.  Each case consists of an ensemble average over 250 configurations.  Top~Left:  Omitting the zero mode.  Top~Right: Omitting modes within the resonance band.  Bottom~Left:  Omitting modes with $k < 4.4$.  Bottom~Right: Omitting modes with $k>2$. }
\label{fig:cuts}
\end{center}
\end{figure}

In order to try to interpret this result and understand the role of different excitations we perform additional calculations using a modified spectrum of fluctuations.  Even though these modified spectra will result in the incorrect quantum expectation values the results may serve useful in understanding the role of different fluctuations.  In Fig.~\ref{fig:cuts} we show the resulting pressure after an ensemble average of 250 configurations for various initial spectra which we now discuss.  In one case (upper left figure) we omit quantum fluctuations of the zero mode.  In a second case (upper right figure) we omit any initial fluctuation within the resonance band.  The lower two figures show spectra which omit any initial fluctuation having $k<4.4$ (lower left) and $k>2$ (lower right).  Of course, once the time evolution begins, there is nothing stopping self interactions from causing excitations to scatter into these initially unoccupied modes.  Not including the resonance modes in the initial spectrum of fluctuations (as done in the top right of Fig.~\ref{fig:cuts}) is different from the analysis of \cite{Dusling:2010rm} where the lattice cutoff was chosen to be below the resonance band.  In the latter case, the resonance modes can never become occupied, which was found to significantly modify the evolution of the pressure.

There is a lot one can infer from Fig.~\ref{fig:cuts}.  The upper two figures comprise results missing a small fraction of the initial spectrum.  In the first case we neglected initial fluctuations of the zero mode while in the second case we are neglecting the very few modes that sit within the narrow resonance band.  In both of these cases the results are qualitatively similar to the results using the full spectrum shown in Fig.~\ref{fig:Full}.  In the case of the zero--mode it is not surprising that its absence doesn't affect the result.  One can estimate the relaxation time from the zero--mode alone based on the previous section where we showed in Eq.~\ref{eq:tauhomo} that the relaxation time in the homogeneous case is inversely proportional to the standard deviation of the Gaussian fluctuations as given by the power spectrum.  In this case
\beqa
\tau_{\rm relax}\approx\frac{2}{g\sqrt{\mathcal{P}_\phi({\bf k}=0)}}\approx 180 
\label{eq:tauZero}
\eeqa
for $g=0.5$.  Clearly, this is a much longer relaxation time then observed in the full 3D simulation.  In the case of the absent resonance modes, one can see by looking at the occupation numbers, that the occupied modes are very quick to scatter and perturb the initially unoccupied resonance band.

What is more interesting is if we neglect a large portion of the initial spectra.  The lower left plot shows the ensemble averaged pressure with a spectra including the intermediate momentum modes (basically we include modes higher than the resonance band).  The evolution from $0\leq t\lsim 50$ is remarkably similar to the result using the full spectrum.  In the lower right plot we have used a spectrum consisting of only low momentum modes.  In this case we no longer have the rapid relaxation at $t \sim 50$ but instead have a gradual relaxation that extends to $t\sim 150$.  Based on this analysis we can understand the two--stage relaxation observed when using the full spectrum.  The first relaxation in the period $0\leq t\lsim 50$ is clearly controlled in some manner by quantum fluctuations above the resonance band.  While the second (more gradual) stage of relaxation, taking place for $t \gsim 50$ is controlled by modes below the resonance band.  The time scale for relaxation due to the low momentum modes is on the order of that estimated in Eq.~\ref{eq:tauZero} for the zero--mode.  Of course this interpretation is only qualitative.  Self--interactions immediately cause modes which are initially unoccupied to become occupied and the result becomes a complex interplay between many modes which cannot be understood simply by studying the linear evolution of individual quanta.

It is interesting to note that the time scale for prethermalization in the three dimensional simulation has the same order of magnitude (it relaxes about 3--4 times faster) as the crude estimate of equation \ref{eq:tauZero} found for the homogeneous case.  If we take the estimate from eq.~\ref{eq:tauZero} seriously we see a faster relaxation with increasing coupling constant $g$.  The power spectrum entering into eq.~\ref{eq:tauZero} is determined by the quantum field theory in question.  Modes with lower $k$ will have the largest fluctuations.  In our case $\mathcal{P}_\phi({\bf k}=0)\sim 1/(g\phi_0)$.  Since $\phi_0\sim 1/g$ the amplitude of the quantum fluctuations are always $\mathcal{O}(1)$.

It is clear that as we vary $g$ the energy density of the system which is of order $g^2 \phi_0^4$ varies as well.  In order to see the parametric behavior of the relaxation time on $g$ at fixed energy density we instead take $\phi_0\sim 1/g^{1/2}$.  In this case we find that
\beqa
\tau_{\rm relax}\sim \frac{1}{g\sqrt{\mathcal{P}_\phi({\bf k}=0)}}\sim \frac{1}{g^{3/4}}
\eeqa
which is consistent with the $1/g^{2/3}$ behavior extracted from the 3D simulation of \cite{Dusling:2010rm}.

\begin{figure}
\begin{center}
\includegraphics[scale=0.8]{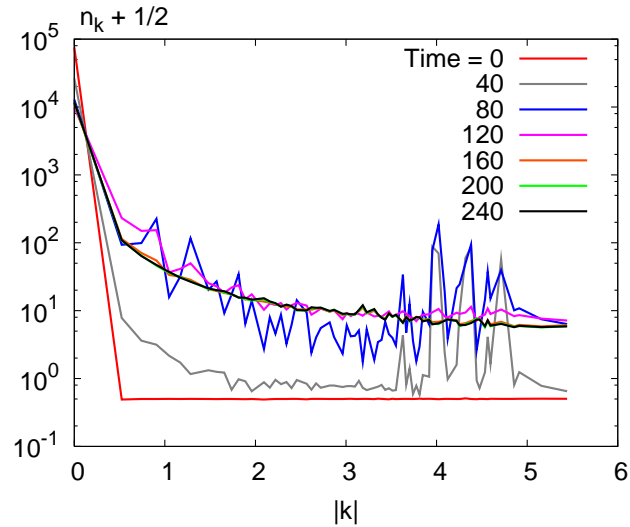}
\caption{Evolution of the number density at various times. The curves at $t=160$ and $t=200$ are almost indistinguishable form the spectra at $t=240$.} 
\label{fig:nk}
\end{center}
\end{figure}

In Fig.~\ref{fig:nk} we show the number density defined by
\beqa
n_{\bf k}\equiv \langle 0\vert \hat{a}^\dagger_{\bf k} \hat{a}_{\bf k}\vert 0 \rangle=\frac{1}{2}\left(\omega_{\bf k} \vert \phi_{\bf k}\vert^2+\frac{\vert \dot{\phi}_{\bf k}\vert^2}{\omega_{\bf k}}\right)-\frac{1}{2}
\eeqa
at various times along the evolution.  The initial condition is such that the number density is zero, $n_{\bf k}=0$, except for the zero--mode which is highly occupied.  As the system evolves one sees the appearance of peaks.  Whether these peaks correspond to resonance modes is not clear.  Even though we know the location of the resonance band at $t\approx 0$ the effective mass of the background field changes with time thereby changing the location of the resonance band with time.  Of course, larger lattice simulations will be needed to reinforce these statements.  At late times, when the system has fully relaxed, the number density is smooth with a power--law fall off.
  
\begin{figure}
\begin{center}
\includegraphics[scale=0.8]{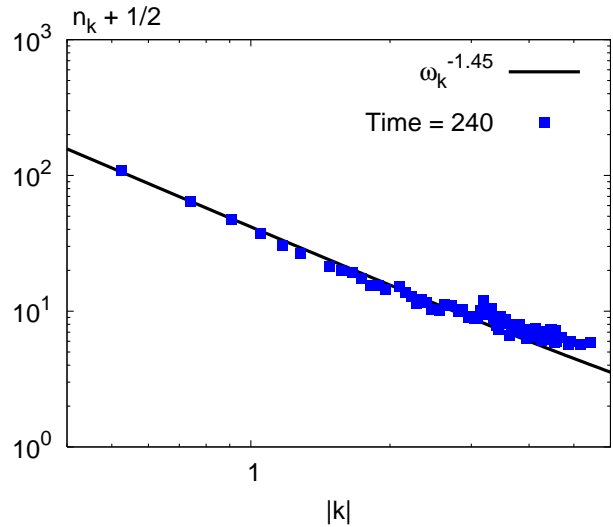}
\caption{Final number density versus momentum.  The solid curve is the best fit to $n_{\bf k}\sim \omega_{\bf k}^{-s}$ with $s=1.45$.}
\label{fig:nkfinal}
\end{center}
\end{figure}

In Fig.~\ref{fig:nkfinal} we show the final number density.  In this case we have plotted the spectra at the discrete values allowed by our grid.  The solid curve is a fit to $n_{\bf k}\sim \omega_{\bf k}^{-s}$ with $s=1.45$.  It clearly does not fall as $1/\omega_{\bf k}$ as one would expect from classical thermal equilibrium.  Interesting further work might study the late time behavior of the particle number to see if it scales according to Kolmogorov turbulence \cite{Micha:2004bv,Arnold:2005qs}.

\section{Conclusions}

In conclusion, it is apparent that quantum fluctuations modify the evolution of a classical scalar theory to the point where it evolves in accordance with ideal hydrodynamics.  We have attempted to understand this behavior by using modified spectra of fluctuations and by studying the particle content of the theory.  We observe that there is a two--stage relaxation process; the rapid early time relaxation is somehow controlled by modes of intermediate momentum (at and above the resonance band) followed by a longer more gradual relaxation which is controlled by the lower momentum (near zero) modes.  The methods used for the scalar field can presumably be extended to the case of gauge theories.

\ack
I would like to thank my collaborators, Thomas Epelbaum, Fran\c ois Gelis and Raju Venugopalan, without whom this work would not have been possible.  This work was funded by US Department of Energy grant DE-FG02-03ER41260.
\newline
\newline

\end{document}